\documentclass[12pt]{iopart}
\usepackage{iopams}  
\usepackage{graphicx}
\usepackage{pifont}

\begin{document}

\title[A stochastic Stirling engine powered by an active particle]{A stochastic Stirling engine powered by an active particle}

\author{Erick Efrain Cote-Valencia and Juan Ruben Gomez-Solano$^{1,*}$}

\address{$^1$Instituto de F\'isica, Universidad Nacional Aut\'onoma de M\'exico, Ciudad de M\'exico, C.P. 04510, M\'exico}
\ead{r\_gomez@fisica.unam.mx}
\vspace{10pt}
\begin{indented}
\item[]May 2025
\end{indented}

\begin{abstract}
We investigate a model of a stochastic engine operating cyclically at constant bath temperature, which consists of an overdamped Brownian harmonic oscillator that plays the role of working substance and is elastically coupled to an active particle. Stirling-like cycles are implemented by
time-periodic changes of the active particle speed and the potential confining the oscillator. By analyzing its quasistatic and finite-time performance, we distinguish two regimes, which are determined by the relative
magnitudes of the persistence time of active particle, the characteristic time of the elastic
interaction, and the viscous relaxation time of the working substance.
In the quasistatic limit, we derive analytic expressions for the average output work,
the mean absorbed heat, and the efficiency, which explicitly involve the ratios of such
three time-scales. For sufficiently large finite cycles, the different quantities characterizing the engine performance converge to their quasistatic values.
For shorter cycles, the efficiencies are systematically lower, but with positive mean output powers that exhibit a maximum as a function of the cycle duration if the corresponding quasistatic mean work is positive. For sufficiently short cycles or if the quasistatic work is zero, the engine behaves at finite-time as a heat pump with negative power and efficiency. 
\end{abstract}

%
% Uncomment for keywords
\vspace{2pc}
\noindent{\it Keywords}: stochastic heat engines, stochastic thermodynamics, active matter, active Brownian engines, nonequilibrium processes
%
% Uncomment for Submitted to journal title message
%\submitto{\JPA}
%
% Uncomment if a separate title page is required
%\maketitle
% 
% For two-column output uncomment the next line and choose [10pt] rather than [12pt] in the \documentclass declaration
%\ioptwocol
%

\section{Introduction}\label{sect:intro}

In recent years, the study of stochastic heat engines has been a subject of major interest in statistical mechanics, as they represent an archetype of non-equilibrium systems that convert the energy extracted from their environment into useful work at mesoscopic scales \cite{holubec2022,fu2023}. Similarly to the pivotal role of their macroscopic counterparts in the development of equilibrium thermodynamics \cite{carnot1824,callen1985}, stochastic heat engines hold potential not only for technological applications \cite{kim2016} but also for shedding light on fundamental properties of fluctuating energy transfers in small systems coupled to heat baths. In fact, based on models of confined Brownian particles as working substances in contact with thermostats periodically driven under various time-varying protocols, numerous aspects of the performance of stochastic heat engines have been investigated in the framework of stochastic thermodynamics, which extends the definitions of work, heat and entropy to the level of stochastic trajectories in such processes \cite{sekimoto1998,seifert2012}. For example, quasistatic and finite-time efficiencies at maximum power and other relevant optimal protocols have been calculated for Carnot \cite{schmiedl2008,tu2014,rana2014,dechant2017,holubec2018,chen2022,frim2022,contrerasvergara2023}, Stirling \cite{contrerasvergara2023,kumari2020,gomezsolano2021,majumdar2022,kumari2023}, Otto \cite{xu2022,xu_2022}, and Ericsson cycles \cite{contrerasvergara2023,kaur2025}. Moreover, optical tweezers experiments on stochastic heat engines like those addressed in the aforementioned theoretical studies have been successfully carried out by using single colloidal beads undergoing various thermodynamic cycles in simple \cite{blickle2012,albay2021,roy2021,krishnamurthy2023,li2024} and complex fluids \cite{roy2023} that act as heat reservoirs. In such experiments, expansions and compressions of the working substance are realized by decreasing and increasing the trap stiffness, respectively, whereas changes in the bath temperature are achieved by an actual variation of the local temperature around the trapped particle or by adding external noise with thermal-like properties. Nowadays, it is well established that in such instances
the mean input heat and the mean output work per cycle have efficiencies that are bounded by the Carnot limit determined by the high and low temperatures of the heat reservoirs \cite{sekimoto2000}, with fluctuations also exhibiting rather general statistical properties \cite{xu2022,xu_2022,lahiri2012,verley2014,manikandan2019,watanabe2022,vanvu2024}.

Inspired by cyclic heat engines consisting of driven Brownian particles that exchange energy with thermal baths, a novel type of mesoscopic heat engines has recently been conceived, in which the working substance or the reservoirs are made of active matter \cite{pietzonka2019,holubec2020,fodor2021}. The components of such active systems, e.g. motile microorganisms and synthetic self-propelled colloids, are capable of autonomously extracting energy from their environment and converting it into directed motion \cite{elgeti2015,bechinger2016}, which can be further exploited as a source of energy for the engine. Therefore, unlike the molecules of a genuine thermal bath, active particles are intrinsically out of equilibrium, which poses theoretical challenges in the description of the performance of stochastic engines periodically operating with active matter. In particular, a pioneering experiment on a colloidal Stirling engine working in a bacterial reservoir with activity tuned by cyclic temperature changes raised the question of whether the quasistatic efficiency of active Brownian engines could exceed the corresponding Carnot limit \cite{krishnamurthy2016}, which has prompted recent efforts to extend the theoretical framework of stochastic thermodynamics to active systems \cite{speck2016,mandal2017,pietzonka2018,dabelow2019,datta2022,bebon2025}. Along these lines, numerous theoretical investigations have been conducted on diverse properties that can affect the performance of active Brownian heat engines, such as periodic changes in friction \cite{saha2018}, and noise persistence \cite{saha2019} from the active bath; activity of the working substance \cite{kumari2020,martin2018,kumari2021,speck2022}; non-Gaussianity \cite{majumdar2022,zakine2017,lee2022}, non-Markovianity  \cite{lee2022,lee2020,guevaravaladez2023,chang2023}, and temperature dependence \cite{kwon2024} of the active noises originating from non-equilibrium reservoirs; memory friction caused by the active bath \cite{guevaravaladez2023,chang2023,holubec__2020}; and underdamped dynamics \cite{holubec__2020}, as well as optimal protocols \cite{martin2018,ekeh2020,szamel2020,gronchi2021}. Some of these issues have also been examined in experiments on active engines realized by colloidal particles subject to either external noise with carefully engineered properties \cite{albay2023} or activation by light absorption \cite{nalupurackal2023}. An important step forward in the description of active heat engines is the identification of conditions for the existence of effective temperatures that can be introduced to map their thermodynamic behavior to one respecting Carnot-like bounds \cite{holubec2020,holubec__2020,gronchi2021,wiese2024,wiese_2024}.

In this paper, we study a model of an active Brownian engine operating periodically in a viscous
fluid at constant temperature. The working substance is composed of an overdamped Brownian particle confined by a harmonic potential, which interacts with an active particle that is part of the environment. Unlike prior investigations using active fluctuating forces acting instantaneously on the engine or short-ranged repulsive potentials, here we consider a harmonic interaction in order to include an additional relaxation time due to the mechanical coupling between the working substance and its surroundings, thus mimicking the engine operation in a viscoelastic-like environment \cite{guevaravaladez2023}. Stirling-like cycles are imposed by means of temporally cyclic variations of the propulsion speed of the active particle and the stiffness of the harmonic potential
confining the working substance. The analysis of the quasistatic and finite-time operation of this model discloses two performance regimes in the parameter space of the system. These are characterized by well-defined features of the quasistatic efficiency, finite-time efficiency, and mean output power, and are determined by the
relative magnitudes of the persistence time of the active particle, the characteristic time of its interaction with the working substance, and the viscous relaxation time of the latter. Our results show that the performance of the active engine investigated here resembles that of conventional stochastic Stirling engines operating in contact with equilibrium thermal baths, and in some special limits it reduces to some instances analyzed in previous studies.

The paper is organized as follows. In \sref{sect:model} we detail the model of the active stochastic engine that was previously described. The results are then presented in \sref{sec:res}, where a derivation of analytic expressions of distinct quantities characterizing the engine performance is provided in the quasistatic limit, while numerical results are discussed in the case of finite cycles. Finally, in \sref{sec:conc} we conclude and provide further remarks.

\section{Model}\label{sect:model}

\begin{figure}[htb]
 \centering
\includegraphics[width=0.9\columnwidth]{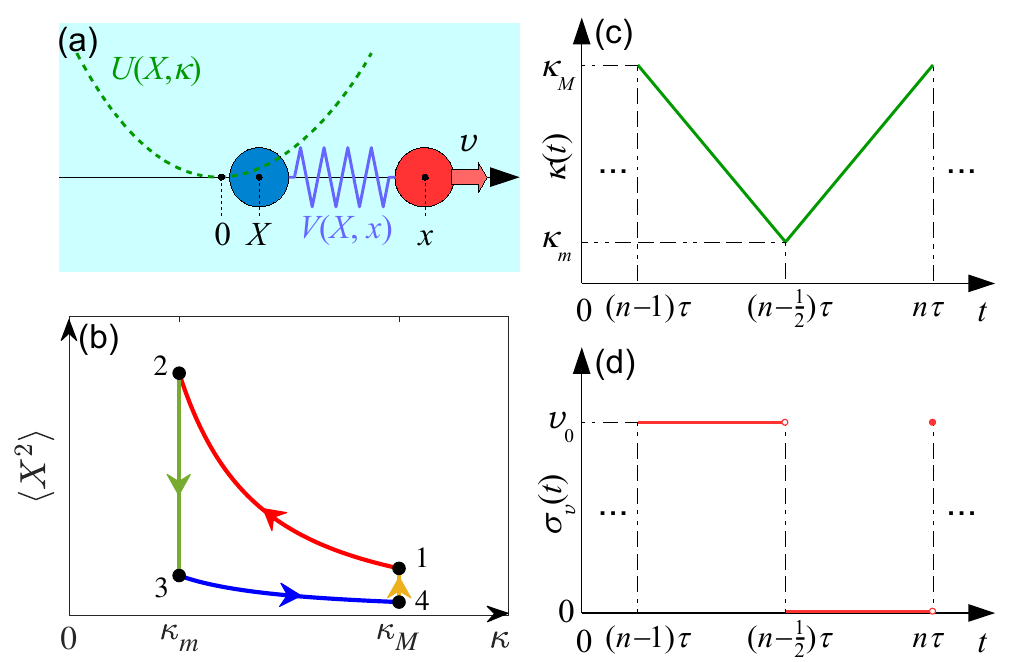}
\caption{(a) Sketch of an isothermal stochastic engine working in a viscous fluid at constant temperature, which is composed of a Brownian particle (blue circle) with position $X$, subject to a confining harmonic potential $U(X,\kappa)$ (green dashed line), and coupled to an active particle (red circle) with position $x$ and propulsion speed $v$ (red arrow) through an interaction potential $V(X,x)$ (blue solid line). (b) Diagram of the four-step Stirling-like cycle of the engine particle, which is characterized by the variance of its position  and the stiffness of the harmonic potential. (c) Temporal variation of the trap stiffness of the confining potential $U(X,\kappa)$ during the $n-$th cycle, as described by \eref{eq:Stirlingkappa}. (d) Temporal variation of the standard deviation of the self-propulsion speed $v$ of the active particle during the $n-$th cycle, as described by \eref{eq:Stirlingspeed}.}
 \label{fig:1}
\end{figure}

We consider a one-dimensional model of a stochastic heat engine isothermally working in a viscous fluid kept at constant temperature $T$, as represented in \fref{fig:1}(a). The engine is composed of a Brownian particle that plays the role of the working substance with a constant friction coefficient $\gamma$, whose position at time $t$ is denoted as $X(t)$. In the following, this particle will be referred to as \emph{engine particle}. The random motion of the engine particle is confined by a harmonic potential, $U(X,\kappa) = \frac{1}{2} \kappa X^2$, whose stiffness $\kappa$ can be controlled externally over time. In addition, the engine particle is elastically coupled through an interaction potential $V(X,x) = \frac{1}{2}k(X-x)^2$ of constant stiffness $k > 0$ to an \emph{active particle}, which is immersed in the same fluid with a friction coefficient $\gamma_A$, where $x(t)$ represents its position at time $t$. The active particle, which will be considered part of the environment with which the engine particle can exchange energy, is not affected by the potential $U(X,\kappa)$ and exhibits a fluctuating self-propulsion speed $v(t)$ characterized by a constant persistence time $\tau_r$ and a standard deviation $\sigma_v(t)$ that can be tuned with time. Accordingly, the stochastic differential equations that describe the overdamped motion of the engine particle and the active particle are
\begin{eqnarray}\label{eq:LangevinBrownianengine}
 \gamma\frac{d}{dt}X(t) & = &-\kappa(t) X(t) - k \left[ X(t) - x(t)\right] + {\zeta}_{X}(t), \nonumber\\
 \gamma_{A}\frac{d}{dt}x(t) & = & - k \left[ x(t) -  X(t) \right] + \gamma_A v(t) + \zeta_x(t),
\end{eqnarray}
where ${\zeta}_{X}(t)$, ${\zeta}_{x}(t)$ and $v(t)$ are Gaussian noises with means
\begin{equation}\label{eq:meannoise}
    \langle \zeta_X(t) \rangle = \langle \zeta_x(t) \rangle = \langle v(t) \rangle = 0,
\end{equation}
cross-correlations
\begin{equation}\label{eq:crosscorrelnoise}
    \langle \zeta_X(t) \zeta_x(t) \rangle = \langle \zeta_X(t)v(t')\rangle  =  \langle \zeta_x(t) v(t') \rangle = 0
\end{equation}
and autocorrelations
\begin{eqnarray}\label{eq:autocorrelnoise}
    \langle {\zeta}_{X}(t) \zeta_X (t') \rangle & = & 2k_B T \gamma \delta(t-t'),\nonumber\\
    \langle {\zeta}_{x}(t) {\zeta}_{x}(t') \rangle & = & 2k_B T \gamma_{A}\delta(t-t'), \nonumber\\
    \langle v(t) v(t')  \rangle & = & \sigma_v(t) \sigma_v(t') \exp \left( - \frac{|t-t'|}{\tau_r}\right).
\end{eqnarray}
Furthermore, we examine  four-step Stirling-like cycles performed by this stochastic engine like the one depicted in \fref{fig:1}(b), during which, in analogy to a gas contained in a piston-cylinder device, the variance of the engine particle position plays the role of the inverse of the pressure, whereas the trap stiffness is similar to the inverse of the gas volume. In \fref{fig:1}(c) we illustrate the temporal change of the trap stiffness during the $n-$th cycle of duration $\tau$ ($n = 1,2,\ldots$), which consists of a piecewise linear dependence of $\kappa(t)$ on time $t$
\begin{equation}\label{eq:Stirlingkappa}
    \kappa(t) = \left\{
    \begin{array}{ll}
    \kappa_M - \frac{2}{\tau} \delta \kappa ~  \left[ t - (n-1)\tau \right], & \,\,\,\,\, (n-1)\tau \le t \le \left( n - \frac{1}{2} \right) \tau,\\
    \kappa_m - \delta \kappa \left\{1-\frac{2}{\tau}\left[t - (n-1)\tau \right] \right\}, & \,\,\,\,\, \left( n - \frac{1}{2} \right) \tau < t \le n \tau,
    \end{array} \right.
\end{equation}
with $\delta \kappa = \kappa_M - \kappa_m > 0$. Moreover, a step-like temporal variation of $\sigma_v(t)$ during the $n-$th cycle is represented in \fref{fig:1}(d), and is described by 
\begin{equation}\label{eq:Stirlingspeed}
    \sigma_v(t) = \left\{
    \begin{array}{ll}
    v_0, & \,\,\,\,\, (n-1)\tau \le t < \left(n - \frac{1}{2}\right)\tau,\\
    0, & \,\,\,\,\, \left(n - \frac{1}{2}\right)\tau \le t < n\tau,\\
   v_0, & \,\,\,\,\,  t = n\tau,
    \end{array} \right.
\end{equation}
where $v_0 > 0$. Note that the temperature changes of the bath in the conventional Stirling cycle are replaced here by changes in the characteristic self-propulsion speed of the active particle, as depicted in \fref{fig:1}(d). More specifically, the $n-$th cycle follows a sequence of four steps:
\begin{itemize}
	\item[$1\rightarrow 2$:]{An expansion with constant activity, $\sigma_v(t) = v_0 > 0$, by means of a linear decrease of $\kappa(t)$ from $\kappa_M$ to $\kappa_m$ during $(n-1)\tau \le t < \left(n-\frac{1}{2} \right)\tau$}.
	\item[$2\rightarrow 3$:]{A sudden decrease at $t = \left(n-\frac{1}{2} \right)\tau$ of $\sigma_v(t)$ from $v_0$ to $0$  with constant $\kappa \left(t = \left(n-\frac{1}{2} \right)\tau \right) = \kappa_m$.}
	\item[$3\rightarrow 4$:]{A compression with zero-activity, $\sigma_v(t) = 0$, through a linear increase of $\kappa(t)$ from $\kappa_m$ to $\kappa_M$ during $\left(n-\frac{1}{2} \right)\tau < t < n\tau$}.
	\item[$4\rightarrow 1$:]{An instantaneous increase at $t = n\tau$ of $\sigma_v(t)$ from $0$ to $v_0$, with $\kappa(t = n\tau) = \kappa_M$, thereby completing the cycle.}
\end{itemize}
We focus on the performance of this stochastic engine after $n \gg 1$ cycles, in which case the effect of the initial conditions $X_0 = X(0)$, $x_0 = x(0)$ must vanish, since the system is expected to achieve a periodic stationary state that is statistically invariant under time translations $t \rightarrow t + \tau$. According to stochastic energetics \cite{sekimoto1998}, in such conditions, the work done {\emph{by the engine particle}} over an infinitesimal time interval $[t,t+dt]$ of duration $dt$, during which the stiffness $\kappa$ changes an amount $d\kappa = \frac{d}{dt}\kappa(t) dt$, is given by
\begin{eqnarray}\label{eq:dW}
    dW & = & -\frac{\partial U(X,\kappa)}{\partial \kappa} \frac{d}{dt} \kappa(t) dt,\nonumber\\
    & = & -\frac{1}{2}X(t)^2 \frac{d}{dt}\kappa(t)dt.
\end{eqnarray}
On the other hand, if $dX$ is the displacement of the engine particle during the same time interval, the resulting change in the harmonic potential energy $U(X,\kappa)$ is
\begin{equation}\label{eq:1stlawengine}
    dU = \frac{\partial U(X,\kappa)}{\partial X} \circ dX + \frac{\partial U(X,\kappa)}{\partial \kappa} \frac{d}{dt}\kappa(t)dt,
\end{equation}
where $\circ$ denotes the Stratonovich product. Note that if the work done by the interaction force on the engine particle, $-\partial_X V \circ dX$, is considered a part of the energy exchanged with the environment (the surrounding fluid plus the active particle), then the first term on the right-hand side of \eref{eq:1stlawengine} can be interpreted as the heat received by the engine particle, $dQ = \left( - \gamma \dot{X} - \partial_X V + \zeta_X \right)\circ dX$. In this way, the energy balance for the engine particle that ignores the degree of freedom $x$ of the active particle can be expressed as
\begin{equation}\label{eq:firstlaw}
    dU = dQ - dW.
\end{equation}
We point out that the complete stochastic energetics of the coupled system composed of the engine particle and the active particle actually reads $dU + dV = dQ' + dQ'_A - dW$ with $dQ' = \left( -\gamma \dot{X} + \zeta_X \right) \circ dX$ and $dQ'_A = \left( -\gamma_A \dot{x} + \zeta_x + \gamma_A v \right) \circ dx$ the amount of heat absorbed from the fluid by the engine and the active particle, respectively, and $dW$ defined in \eref{eq:dW}. Then, the heat $dQ$ considered here satisfies $dQ + dV = dQ' + dQ'_A$, i.e. the total heat absorbed from the fluid by both particles is partitioned between the work done on the engine particle by all of the forces of the environment acting on it and the change in the interaction potential.

According to \eref{eq:dW}, the work done on average by the particle during the time interval $(n-1) \tau \le t \le n\tau$, i.e. during the full $n-$th cycle, is 
\begin{eqnarray}\label{eq:meanwork}
    \langle W_{\tau} \rangle & = & -\frac{1}{2}\int_{(n-1)\tau}^{n\tau} dt \, \frac{d\kappa}{dt} \langle X(t)^2 \rangle, \nonumber\\
    & = & \frac{\delta \kappa}{\tau} \left[ \int_{(n-1)\tau}^{\left(n - \frac{1}{2}\right)\tau} dt\, \langle X(t)^2 \rangle - \int_{\left(n - \frac{1}{2}\right)\tau}^{n\tau} dt \, \langle X(t)^2 \rangle  \right],
\end{eqnarray}
where we have made use of the temporal change of $\kappa(t)$ given by \eref{eq:Stirlingkappa}, and $\langle \ldots \rangle$ represents an average with respect to the noises $\zeta_X$, $\zeta_x$, and $v$.
Note that \eref{eq:meanwork} results from the sum of the works performed on average during each of the four steps of the cycle, where the contributions of the steps $2 \rightarrow 3$ and $4 \rightarrow 1$ vanish because $\kappa$ remains constant. Likewise, \eref{eq:firstlaw} leads to the following expression for the mean energy transferred as heat to the engine particle from its environment over the half of the $n-$th cycle during which the self-propulsion of the active particle is switched on from $\sigma_v(t) = 0$ to $\sigma_v(t) = v_0$, i.e. during steps $4 \rightarrow 1$ of the $(n-1)-$th cycle and $1 \rightarrow 2$ of the $n-$th cycle
\begin{eqnarray}\label{eq:meanheat}
    \langle Q_{\tau/2} \rangle   & = & \frac{1}{2} \left[ \kappa_m \left\langle X \left(\left(n - \frac{_1}{^2}\right)\tau^- \right)^2 \right\rangle - \kappa_M \left\langle X((n-1)\tau^-)^2 \right\rangle \right] \nonumber\\ 
    & & + \frac{\delta \kappa}{\tau}  \int_{(n-1)\tau}^{\left(n - \frac{1}{2}\right)\tau} dt\, \langle X(t)^2 \rangle.
\end{eqnarray}
Then, the efficiency of this engine during a complete Stirling-like cycle of duration $\tau$ can be defined as
\begin{equation}\label{eq:efficiency}
    \mathcal{E}_{\tau} = \frac{\langle W_{\tau} \rangle}{\langle Q_{\tau/2} \rangle}.
\end{equation}

\section{Results}\label{sec:res}

\subsection{Quasistatic performance}\label{subsect:quasistatic}

We proceed to investigate the quasistatic behavior of the engine, i.e. when the Stirling-like cycle occurs infinitely slowly and the effects of the initial conditions vanish. In such a case, the term $\langle X(t)^2 \rangle$ of the integrands of \eref{eq:meanwork} and \eref{eq:meanheat} is related to the non-equilibrium steady-state variance of $X$ that does not explicitly depend on $t$ but on the instantaneous values of $\kappa(t)$ and $\sigma_v(t)$. To find an analytic expression of $\langle X(t)^2 \rangle$ valid in the quasistatic limit $\tau \rightarrow \infty$ and $n \gg 1$, we first find the solution $X(t)$ at time $t > 0$ of \eref{eq:LangevinBrownianengine} with constant $\kappa(t) = \kappa$ and $\sigma_v(t) = \sigma_v$, noise properties given by \eref{eq:meannoise}-\eref{eq:autocorrelnoise}, and initial conditions $X_0 = X(0)$, $x_0 = x(0)$, which reads
\begin{eqnarray}\label{eq:solX}
    X(t) & = & \gamma X_0 J(t) + \int_0^t dt' \, J(t-t')\zeta_X(t')  \nonumber\\
    & & + \gamma_A x_0 I(t)  + \gamma_A \int_0^t dt'\, I(t-t')v(t') + \int_0^t dt' \, I(t-t')\zeta_x(t'). 
\end{eqnarray}
In \eref{eq:solX}, the time-dependent functions $I(t)$ and $J(t)$ are defined through their Laplace transforms at the complex frequency $s$, $\tilde{I}(s) = \int_0^{\infty} dt\, e^{-st} I(t)$ and $\tilde{J}(s) = \int_0^{\infty} dt\, e^{-st} J(t)$, by the following equations
\begin{equation}\label{eq:LapI}
    \tilde{I}(s) = \frac{k}{\gamma \gamma_A s^2 + \left[k \gamma + (\kappa + k) \gamma_A \right] s + k \kappa}
\end{equation}
\begin{equation}\label{eq:LapJ}
    \tilde{J}(s) = \frac{\gamma_A s + k}{\gamma \gamma_A s^2 + \left[k \gamma + (\kappa + k) \gamma_A \right] s + k \kappa}.
\end{equation}
Upon Laplace inversion of the functions given in \eref{eq:LapI} and \eref{eq:LapJ}, the following explicit expressions for $I(t)$ and $J(t)$ can be derived
\begin{equation}\label{eq:I}
    I(t) = \frac{k}{\gamma \gamma_A}\left( A_+e^{-\epsilon_+ t} + A_- e^{-\epsilon_- t}\right),
\end{equation}
\begin{equation}\label{eq:J}
    J(t) = \frac{1}{\gamma}\left( B_+e^{-\epsilon_+ t} + B_- e^{-\epsilon_- t}\right),
\end{equation}
where $\epsilon_+$, $\epsilon_-$ are two real-valued decay rates of the system given by
\begin{equation}\label{eq.rates}
    \epsilon_{\pm} = \frac{1}{2} \left[\left( \frac{k}{\gamma_A} + \frac{\kappa + k}{\gamma}\right) \pm \sqrt{\left( \frac{k}{\gamma_A} + \frac{\kappa + k}{\gamma}\right)^2-\frac{4\kappa k}{\gamma \gamma_A}}\right],
\end{equation}
which satisfy $\epsilon_+ \ge \epsilon_- \ge 0$, and $A_+$, $A_-$, $B_+$, $B_-$ are coefficients defined by the expressions
\begin{eqnarray}\label{eq:coeffAB}
    A_{\pm} & = & \mp \left[ \left( \frac{k}{\gamma_A} + \frac{\kappa + k}{\gamma}\right)^2-\frac{4\kappa k}{\gamma \gamma_A} \right]^{-1/2},\nonumber\\
    B_{\pm} & = & \frac{1}{2}\left[ \mp \left[\left( \frac{k}{\gamma_A} + \frac{\kappa + k}{\gamma}\right)^2-\frac{4\kappa k}{\gamma \gamma_A}\right]^{-1/2}\left( \frac{k}{\gamma_A} - \frac{\kappa + k}{\gamma}\right) + 1\right].
\end{eqnarray}
Under such conditions, \eref{eq:solX} leads to the average position of the engine particle at time $t > 0$ 
\begin{equation}\label{eq:meanX}
    \langle X(t) \rangle = \gamma X_0 J(t) + \gamma_A x_0 I(t).
\end{equation}
Therefore, using \eref{eq:solX} and \eref{eq:meanX} along with the noise properties described by \eref{eq:meannoise}-\eref{eq:autocorrelnoise}, the variance of the position of the engine particle at time $t > 0$ can be expressed as 
\begin{eqnarray}\label{eq:varX}
    \sigma_X(t)^2 & \equiv & \langle \left[ X(t) - \langle X(t) \rangle \right]^2 \rangle \nonumber\\
     & = & \int_0^t dt' \int_0^t dt'' I(t-t') I(t-t'') \left[ \gamma_A^2 \langle v(t') v(t'') \rangle + \langle \zeta_x(t') \zeta_x(t'') \rangle \right]\nonumber\\\
    & & + \int_0^t dt' \int_0^t dt'' J(t-t') J(t-t'') \langle \zeta_X(t') \zeta_X(t'') \rangle.
\end{eqnarray}
It should be noted that, for $t \rightarrow \infty$, $\sigma_X(t)^2$ reaches a steady-state value with $\langle X(t) \rangle \rightarrow 0$, i.e. the effect of the initial conditions $X_0$ and $x_0$ become completely negligible because of the exponentially decaying behavior of the functions $I(t)$ and $J(t)$ described by \eref{eq:I} and \eref{eq:J}. Consequently,  the quasistatic value of $\langle X(t)^2 \rangle$ needed for the calculation of $\langle W_{\tau} \rangle$ and $\langle Q_{\tau/2 }\rangle$ in the limit $\tau \rightarrow \infty$ reduces to the steady-state value of $\sigma_X(t)^2$. Therefore, by substituting the noise autocorrelation functions of \eref{eq:autocorrelnoise} into \eref{eq:varX} and taking the limit $t \rightarrow \infty$, the following expression for $ \langle X^2 \rangle \equiv \langle X(t \rightarrow \infty)^2 \rangle$ can be found
\begin{eqnarray}\label{eq:varXstat}
    \langle X^2 \rangle & = & \lim_{t \rightarrow \infty} \sigma_X(t)^2,\nonumber\\
    & = & \frac{k_B T}{\kappa} + \frac{\left(1 + \frac{ \gamma}{\gamma_A +\gamma + \tau_A \kappa} \frac{\tau_A}{\tau_r}\right) \gamma_A^2 \sigma_v^2}{\left[ \frac{1}{\tau_r} \left(\gamma_A + \gamma + \gamma \frac{\tau_A}{\tau_r}  \right) + \kappa\left( 1 + \frac{ \tau_A}{\tau_r} \right)\right]\kappa}
\end{eqnarray}
where $\tau_A = \frac{\gamma_A}{k}$ represents a viscoelastic-like relaxation time of the active particle due to its interaction with the engine particle. Note that for $\sigma_v = 0$, Eq. (\ref{eq:varXstat}) is consistent with the equipartition relation $\frac{1}{2} \kappa \langle X^2 \rangle = \frac{1}{2}k_B T$ at thermal equilibrium, whereas for $\sigma_v = v_0 > 0$, the equipartition value is increased by an amount proportional to $v_0^2$ and depending nonlinearly on $\kappa^{-1}$ and the time-scales $\tau_r$ and $\tau_A$.

By substituting \eref{eq:varXstat} into \eref{eq:meanwork} and after integrating with respect to $\kappa$ over the two halves of the cycle with the corresponding values of $\sigma_v$, we find the following expression for the mean quasistatic work performed by the engine (normalized by $k_B T$)
\begin{equation}\label{eq:meanWork}
    \frac{1}{k_B T}   \langle W_{\tau \rightarrow \infty} \rangle = \frac{1}{2\delta} \ln \left[ \theta \left( \frac{1 + \nu \epsilon}{1 + \frac{\nu}{\theta} \epsilon}\right)^{\frac{\alpha \epsilon^2}{1 - \alpha \epsilon^2}} \left(\frac{\frac{\nu}{\theta} + \frac{1 + \alpha \epsilon}{1 + \epsilon}}{\nu + \frac{1 + \alpha \epsilon}{1 + \epsilon}} \right)^{\frac{1}{1 - \alpha \epsilon^2}}\right].
\end{equation}
%\begin{equation}\label{eq:meanWork}
%    \langle W_{\tau \rightarrow \infty} \rangle  =  \frac{\gamma_A^2 v_0^2 \tau_r}{2 \gamma_0} \ln \left[ \frac{\kappa_M}{\kappa_m} \left(  \frac{\kappa_M + \frac{\gamma_0}{\tau_A}}{\kappa_m + \frac{\gamma_0}{\tau_A}}\right)^{\frac{\gamma \tau_A^2}{\gamma_0 \tau_r^2 - \gamma \tau_A^2}} \left( \frac{\kappa_m + \frac{\gamma_0 + \gamma \frac{\tau_A}{\tau_r}}{\tau_A + \tau_r}}{\kappa_M + \frac{\gamma_0 + \gamma \frac{\tau_A}{\tau_r}}{\tau_A + \tau_r}} \right)^{\frac{\gamma_0 \tau_r^2}{\gamma_0 \tau_r^2 - \gamma \tau_A^2}}\right],
%\end{equation}
In a similar manner, upon substitution of \eref{eq:varXstat} in \eref{eq:meanheat}, we derive the following formula for the mean heat quasistatically absorbed by the engine particle during the activation of the environment (normalized by $k_B T$)
\begin{eqnarray}\label{eq:meanHeat}
   \frac{1}{k_B T}\langle Q_{\tau/2 \rightarrow \infty} \rangle & = & \frac{1}{2\delta} \frac{1 + \left(\alpha + \frac{\nu}{\theta} \right) \epsilon}{\left(1 + \frac{\nu}{\theta} \epsilon \right) \left[1 + \frac{\nu}{\theta} + \left(\alpha + \frac{\nu}{\theta} \right) \epsilon\right]} \nonumber\\
    & & + \frac{1}{2\delta} \ln \left[ \theta^{1 + \delta} \left( \frac{1 + \nu \epsilon}{1 + \frac{\nu}{\theta} \epsilon}\right)^{\frac{\alpha \epsilon^2}{1 - \alpha \epsilon^2}} \left(\frac{\frac{\nu}{\theta} + \frac{1 + \alpha \epsilon}{1 + \epsilon}}{\nu + \frac{1 + \alpha \epsilon}{1 + \epsilon}} \right)^{\frac{1}{1 - \alpha \epsilon^2}}\right].
\end{eqnarray}
%\begin{eqnarray}\label{eq:meanHeat}
%    \langle Q_{\tau/2 \rightarrow \infty} \rangle & = & \frac{1}{2} k_B T \ln \left( \frac{\kappa_M}{\kappa_m}\right) \nonumber\\
%    & & + \frac{\gamma_A^2 v_0^2 \tau_r}{2 \gamma_0} \ln \left[ \frac{\kappa_M}{\kappa_m} \left(  \frac{\kappa_M + \frac{\gamma_0}{\tau_A}}{\kappa_m + \frac{\gamma_0}{\tau_A}}\right)^{\frac{\gamma \tau_A^2}{\gamma_0 \tau_r^2 - \gamma \tau_A^2}} \left( \frac{\kappa_m + \frac{\gamma_0 + \gamma \frac{\tau_A}{\tau_r}}{\tau_A + \tau_r}}{\kappa_M + \frac{\gamma_0 + \gamma \frac{\tau_A}{\tau_r}}{\tau_A + \tau_r}} \right)^{\frac{\gamma_0 \tau_r^2}{\gamma_0 \tau_r^2 - \gamma \tau_A^2}}\right] \nonumber\\
%    & & + \frac{\gamma_A^2 v_0^2 \tau_r}{2\gamma_0} \frac{1 + \frac{\gamma}{\gamma_0 + \tau_A \kappa_m} \frac{\tau_A}{\tau_r}}{1 + \frac{\gamma \tau_A}{\gamma_0 \tau_r} + \frac{\kappa_m}{\gamma_0} (\tau_A + \tau_r)}.
%\end{eqnarray}
In \eref{eq:meanWork} and \eref{eq:meanHeat}, $\epsilon$ and $\nu$ are dimensionless parameters that define the following ratios of time-scales
\begin{equation}\label{eq:epsilon}
    \epsilon  =  \frac{\tau_A}{\tau_r},
\end{equation}
and
\begin{equation}\label{eq:nu}
     \nu  =  \frac{\tau_r}{\tau_{\kappa_M}},
\end{equation}
respectively, where $\tau_{\kappa_M} = ( \gamma + \gamma_A )/ \kappa_M$ corresponds the fastest viscous relaxation time of the system. Moreover, the parameter $\alpha$ is given by 
\begin{equation}\label{eq:alfa}
    \alpha  =  \frac{\gamma}{\gamma + \gamma_A},
\end{equation}
and represents the ratio of the viscous friction coefficient experienced by the engine particle alone to the the total friction coefficient emerging from its interaction with the active particle, whose possible values are bounded to $0 < \alpha < 1$. In addition, the parameter $\theta$ is 
\begin{equation}\label{eq:theta}
     \theta  = \frac{\kappa_M}{\kappa_m},
\end{equation}
thus playing the role of a compression coefficient, in analogy to a macroscopic Stirling engine working with a gas. Finally, the parameter 
\begin{equation}
     \delta  =  \frac{k_B T}{\gamma_A (1-\alpha) v_0^2 \tau_r},
\end{equation}
quantifies the relative magnitude of thermal fluctuations with respect to the active ones.

\begin{figure}[htb]
 \centering
\includegraphics[width=0.95\columnwidth]{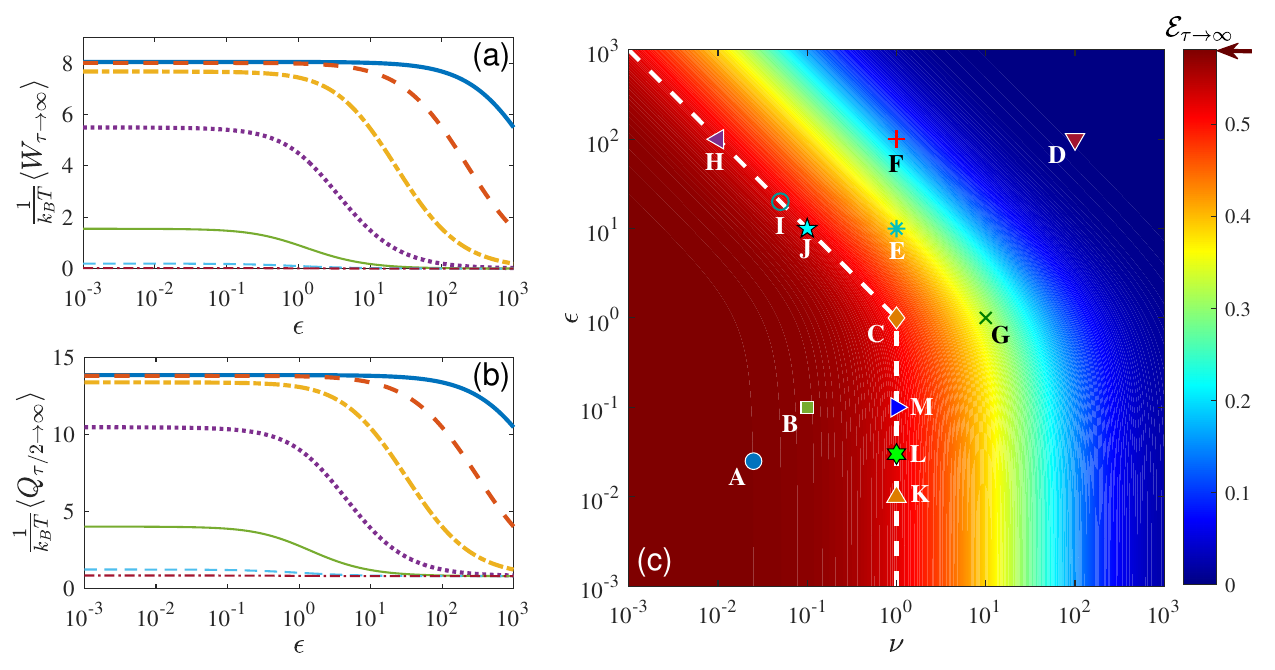}
\caption{[(a) and (b)] Dependence on the parameter $\epsilon$ of: (a) the mean quasistatic work done by the engine; (b) the mean quasistatic heat absorbed from the environment, for distinct values of the parameter $\nu$. From top to bottom: $\nu = 10^{-3}$ (thick solid line), $10^{-2}$ (thick dashed line), $10^{-1}$ (thick dotted-dashed line), $10^0$ (thick dotted line), $10^1$ (thin solid line), $10^2$ (thin dashed line), and $10^3$ (thin dotted dashed line). (c) Quasistatic efficiency of the Stirling engine calculated from \eref{eq:quasistaticefficiency} as a function of $\nu$ and $\epsilon$ for $\alpha = 0.5$, $\theta = 5$ and $\delta = 0.1$. The arrow on the top-right corner of the colorbar indicates the maximum efficiency $\mathcal{E}_{\tau \rightarrow \infty}^{\mathrm{max}} = 0.5809$ that can be achieved for such values of $\theta$ and $\delta$, see \eref{eq:maxquaseff}. The dashed line represents the curve $\nu = 1$ if $0 < \epsilon < 1$ and $\nu = 1/\epsilon$ if $\epsilon \ge 1$ that separates the optimal from the suboptimal performance regimes. The symbols labeled from {\bf A} to {\bf M} depict the pairs of values $(\nu ,\epsilon)$ used for the corresponding finite-cycle simulations.}
 \label{fig:2}
\end{figure}

In \fref{fig:2}(a) and \fref{fig:2}(b) we illustrate the dependence on the parameter $\epsilon$ defined in \eref{eq:epsilon} of the mean work quasistatically delivered per cycle by the engine and the corresponding average heat absorbed from its environment. These were calculated by use of \eref{eq:meanWork} and \eref{eq:meanHeat}, respectively, for exemplary values spanning six orders of magnitude of the parameter $\nu$ defined in \eref{eq:nu} and fixed $\alpha = 0.5$, $\theta = 5$, and $\delta = 0.1$. It is noteworthy that both $\langle W_{\tau \rightarrow \infty} \rangle$ and $\langle Q_{\tau/2 \rightarrow \infty} \rangle$ are monotonically decreasing functions of $\epsilon$ regardless of the specific value of $\nu$. First, with decreasing $\epsilon$ and fixed $\nu$, the mean work saturates to a non-negative value given by
\begin{equation}\label{eq:workconstantnu}
   \frac{1}{k_B T} \lim_{\epsilon \rightarrow 0}  \langle W_{\tau \rightarrow \infty} \rangle = \frac{1}{2\delta} \ln \left( \frac{\nu + \theta}{\nu + 1}  \right),
\end{equation}
and the mean heat reaches a positive value determined by 
\begin{equation}\label{eq:heatconstantnu}
    \frac{1}{k_B T} \lim_{\epsilon \rightarrow 0} \langle Q_{\tau / 2 \rightarrow \infty} \rangle = \frac{1}{2} \ln \theta + \frac{1}{2 \delta} \left[  \ln \left( \frac{\nu + \theta}{\nu + 1} \right) + \frac{\theta}{\nu + \theta} \right],
\end{equation}
irrespective of $\alpha$, as verified by the existence of plateaus at sufficiently small $\epsilon \ll 1$ in \fref{fig:2}(a) and \fref{fig:2}(b). The general trend of such plateau values is that, with increasing $\nu$, they monotonically decline from their absolute maxima 
\begin{eqnarray}\label{eq:maxworkheat}
    \lim_{\nu \rightarrow 0,\epsilon \rightarrow 0} \frac{1}{k_B T} \langle W_{\tau \rightarrow \infty} \rangle & = & \frac{1}{2\delta} \ln \theta, \nonumber\\
    \lim_{\nu \rightarrow 0,\epsilon \rightarrow 0} \frac{1}{k_B T} \langle Q_{\tau / 2 \rightarrow \infty} \rangle & = & \frac{1}{2\delta} \left( \ln \theta^{1+\delta} + 1 \right), 
\end{eqnarray}
to their absolute minima
\begin{eqnarray}\label{eq:minworkheat}
    \lim_{\nu \rightarrow \infty,\epsilon \rightarrow 0} \frac{1}{k_B T} \langle W_{\tau \rightarrow \infty} \rangle & = & 0, \nonumber\\
    \lim_{\nu \rightarrow \infty,\epsilon \rightarrow 0} \frac{1}{k_B T} \langle Q_{\tau / 2 \rightarrow \infty} \rangle & = & \frac{1}{2} \ln \theta,
\end{eqnarray}
respectively. This behavior reveals that an optimal condition for the engine to quasistatically produce on average the maximum amount of work at fixed $\theta$ and $\delta$ is to operate such that $\tau_A \ll \tau_r \ll \tau_{\kappa_M}$, i.e. when the coupling between both particles is very strong ($k \gg \gamma_A /\tau_r$) and the propulsion speed of the active particle decorrelates sufficiently fast. In contrast, when $\tau_A \ll \tau_r$ but $\tau_r \gg \tau_{\kappa_M}$, the propulsion force $\gamma_A v$ acting on the active particle is no longer a stochastic source of energy for the engine particle as this term behaves as a quasi-constant force with infinite persistence. Consequently, even if a non-zero amount of heat is extracted on average from the isothermal environment, the engine is not able to convert it into work. Moreover, when $\epsilon \gg \nu^{-1}$, that is for a very weak particle coupling such that $\tau_A \gg \tau_{\kappa_M}$, the mean work tends to 0 for all possible values of $\nu$, as shown in \fref{fig:2}(a). Furthermore, as shown in \fref{fig:2}(b) the mean heat converges to the positive value $k_B T (\ln \theta)/2$, i.e. the same as in \eref{eq:minworkheat}, which represents the minimum average amount of heat that can be isothermally absorbed from the bath at constant temperature by a quasistatic change $\theta > 1$ of the stiffness $\kappa$. These findings suggest that, unlike the conventional Stirling engine, the quasistatic operation of this isothermal stochastic engine fueled by an active particle exhibits a rather intricate behavior resulting from the interplay between the different time-scales of the system, which will be further analyzed through its efficiency.

Using \eref{eq:efficiency}, \eref{eq:meanWork} and \eref{eq:meanHeat}, we can find the following analytic expression for the quasistatic efficiency of the cycle \begin{equation}\label{eq:quasistaticefficiency}
	\mathcal{E}_{\tau \rightarrow \infty} = \frac{\ln \left[ \theta \left( \frac{1+\nu \epsilon}{1+\frac{\nu}{\theta}\epsilon} \right)^{\frac{\alpha \epsilon^2}{1-\alpha \epsilon^2}} \left( \frac{\frac{\nu}{\theta}+\frac{1+\alpha \epsilon}{1+\epsilon}}{\nu +\frac{1+\alpha \epsilon}{1+\epsilon}} \right)^{\frac{1}{1-\alpha \epsilon^2}}\right]}{\frac{1 + \left( \alpha + \frac{\nu}{\theta}\right)\epsilon}{\left(1 + \frac{\nu}{\theta} \epsilon \right) \left[ 1 + \frac{\nu}{\theta} + \left( \alpha + \frac{\nu}{\theta}\right)\epsilon \right]} + \ln \left[  \theta^{1+\delta} \left( \frac{1+\nu \epsilon}{1+\frac{\nu}{\theta}\epsilon} \right)^{\frac{\alpha \epsilon^2}{1-\alpha \epsilon^2}} \left( \frac{\frac{\nu}{\theta}+\frac{1+\alpha \epsilon}{1+\epsilon}}{\nu+\frac{1+\alpha \epsilon}{1+\epsilon}} \right)^{\frac{1}{1-\alpha \epsilon^2}} \right]}.
\end{equation}
This general expression of the quasistatic efficiency reveals some relevant limiting behaviors of the engine under some particular conditions. For example, when the interaction relaxation time $\tau_A$ is much smaller than the correlation time $\tau_r$
of the self-propulsion speed such that $\epsilon \rightarrow 0$, then the quasistatic efficiency takes the form
\begin{equation}\label{eq:quasieffepsilon0}
    \mathcal{E}_{\tau \rightarrow \infty} = \frac{\ln \left( \frac{\theta + \nu}{1 + \nu} \right) }{\frac{\theta}{\theta + \nu} + \ln \left( \frac{\theta + \nu}{1 + \nu}\right) + \ln \theta^{\delta}}.
\end{equation}
This expression results from taking the ratio between \eref{eq:workconstantnu} and \eref{eq:heatconstantnu} and corresponds to the situation in which the interaction between the engine and the active particle is instantaneous. Therefore, the activity of the latter is immediately transmitted to the former, thus effectively behaving as a stochastic Stirling engine subject to active Ornstein-Uhlenbeck noise. In fact, in the special case where $\delta = 0$, \eref{eq:quasieffepsilon0} reduces to the expression derived in \cite{zakine2017} for this type of engine in the absence of thermal fluctuations. 

A second relevant behavior occurs in the opposite situation when $\tau_A \gg \tau_r$. If $\tau_r$ is nonzero and $\tau_A \rightarrow \infty$ in such a way that $\epsilon \rightarrow \infty$ and $0 < \nu < \infty$, the interaction with the active particle is so weak that the engine particle does not produce work on average, resulting in a vanishing quasistatic efficiency $\mathcal{E}_{\tau \rightarrow \infty} = \ln (\theta \theta^{-1}) / \ln(\theta^{-1} \theta^{1+\delta}) = 0$. A similar situation takes place when $ \tau_A$ remains finite but $\tau_r \rightarrow 0$, in which case $\delta \rightarrow \infty$, $\epsilon \rightarrow \infty$, $\nu \rightarrow 0$, with $\nu \epsilon = \tau_A / \tau_{\kappa_M} < \infty$, and consequently the quasistatic efficiency is also $ \mathcal{E}_{\tau \rightarrow \infty} = 0$. In this scenario, the persistence length of the active particle during the first half of the cycle is $v_0 \tau_r \rightarrow 0$ even if $v_0 > 0$, thus acting effectively as a passive particle throughout the entire cycle from which no usable heat can be converted into work on average at constant temperature of the surrounding fluid.

A third interesting behavior
corresponds to the case where the friction coefficient of the active particle is much larger than that of the engine particle, $\gamma_A \gg \gamma$, which could be possible if, e.g. the size of the former is much larger than that of the latter. In this case, $\alpha \rightarrow 0$, and therefore the quasistatic efficiency becomes
\begin{equation}\label{eq:quasieffgammaAinf}
    \mathcal{E}_{\tau \rightarrow \infty} = \frac{\ln \left[ \frac{\theta + (1 + \epsilon) \nu}{ 1 + (1 + \epsilon) \nu } \right]}{\frac{\theta}{\theta + (1 + \epsilon) \nu} + \ln \left[ \frac{\theta + (1 + \epsilon) \nu}{ 1 + (1 + \epsilon) \nu } \right] + \ln \theta^{\delta}},
\end{equation}
which has the same form as \eref{eq:quasieffepsilon0}, with $\nu$ replaced by $(1 + \epsilon) \nu$, and can be interpreted as the effective motion of the engine particle in a potential with a trap stiffness increased by a factor $1 + \epsilon > 1$ and subject to active Ornstein-Uhlenbeck noise. On the other hand, if $\gamma_A \ll \gamma$, i.e. $\alpha \rightarrow 1$, $\delta \rightarrow \infty$, the quasistatic efficiency vanishes, as the effect of the active particle is negligible with respect to the thermal noise at constant temperature.

It should also be noted that \eref{eq:quasistaticefficiency} is similar to the expression derived for the quasistatic efficiency of a Stirling engine operating in a viscoelastic bath composed of $N \gg 1$ active particles with tunable activity immersed in a fluid at constant temperature \cite{guevaravaladez2023}. In fact, at constant values of the parameters $\alpha$, $\theta$, and $\delta$, equation \eref{eq:quasistaticefficiency} exhibits a similar dependence on $\epsilon$ and $\nu$, as illustrated by the colormap in Fig. \ref{fig:2}(c) for $\alpha = 0.5$, $\theta = 5$, and $\delta = 0.1$, where two qualitatively different regimes describing the performance of the engine can be identified: 

\begin{itemize}
\item{The first, which will be called \emph{optimal regime}, corresponds to the region defined by $0 < \nu < 1$ and $0 < \epsilon  < 1/\nu$ , over which the quasistatic efficiency is almost constant and very close to the following expression obtained from \eref{eq:efficiency} and \eref{eq:maxworkheat}
\begin{equation}\label{eq:maxquaseff}
    \mathcal{E}_{\tau \rightarrow \infty}^{\mathrm{max}} = \frac{1}{1 + \delta + \frac{1}{\ln \theta}}.
\end{equation}
We point out that when $\epsilon \rightarrow 0$ and $\nu \rightarrow 0$, and at fixed $\delta$, it can be easily verified that $\mathcal{E}_{\tau \rightarrow \infty}$ reaches a maximum regardless of the specific value of $\alpha$, which is precisely given by \eref{eq:maxquaseff}. For instance, for the particular example considered in Fig. \ref{fig:2}(c), $\mathcal{E}_{\tau \rightarrow \infty}^{\mathrm{max}} = 0.5809$. Therefore, \eref{eq:maxquaseff} represents an upper bound for the quasistatic efficiency of the Stirling-like engine powered by the active particle, i.e. $\mathcal{E}_{\tau \rightarrow \infty} \le \mathcal{E}_{\tau \rightarrow \infty}^{\mathrm{max}}$. This maximum efficiency is equivalent to that of a stochastic Stirling engine subject to changes in the actual bath temperature, $\left[ \left(1 - \frac{T_c}{T_h}\right)^{-1} + \frac{1}{\ln \theta} \right]^{-1}$, where the temperature of the cold reservoir is kept at $T_c = T$ while the temperature of the hot reservoir is $T_h = \left(1 + \delta^{-1} \right) T$.}
\item{The second regime, which will be called \emph{suboptimal regime},  is defined by the region $\nu \ge 1/\epsilon$ if $\epsilon \ge 1$ and $\nu \ge 1$ if $0 < \epsilon < 1$, where the quasistatic efficiency drops from values below the maximum given by \eref{eq:maxquaseff} to 0. Under such conditions, the engine cannot convert the absorbed heat into output work in a totally optimal manner. In particular, when $\nu \gg 1/\epsilon$ if $\epsilon \ge 1$ or $\nu \gg 1$ if $0 < \epsilon < 1$, the efficiency vanishes because the engine cannot produce work on average for this selection of the parameters $\nu$ and $\epsilon$ independently of the values of $\theta$, $\alpha$ and $\delta$. 
Note that the two regions in the plane $(\nu,\epsilon)$ that define the aforementioned performance regimes are clearly outlined by the curve $\nu = 1$ if $0 < \epsilon < 1$ and $\nu = 1/\epsilon$ if $\epsilon \ge 1$, see dashed line in  \fref{fig:2}(a).}
\end{itemize}

It must be pointed out that the existence of such two regimes in the performance of the engine actually results from the interplay between the three characteristic time-scales of the system. In fact, the optimal conditions for the quasistatic operation at maximum efficiency that occurs in the first regime, namely, $0  < \nu < 1$ and $0 < \epsilon  < 1/\nu$, require that both $\tau_r < \tau_{\kappa_M}$ and $\tau_A < \tau_{\kappa_M}$ are simultaneously satisfied, highlighting the significance of the viscous relaxation time $\tau_{\kappa_M}$ in the engine performance. When these conditions are not met, the energy injected into the engine particle through the softening and stiffening of the confining potential and the coupling with the active particle is dissipated by viscous friction into the surroundings before it is fully harnessed to produce work, thereby resulting in a suboptimal energy conversion.

\begin{figure}[htb]
 \centering
\includegraphics[width=0.95\columnwidth]{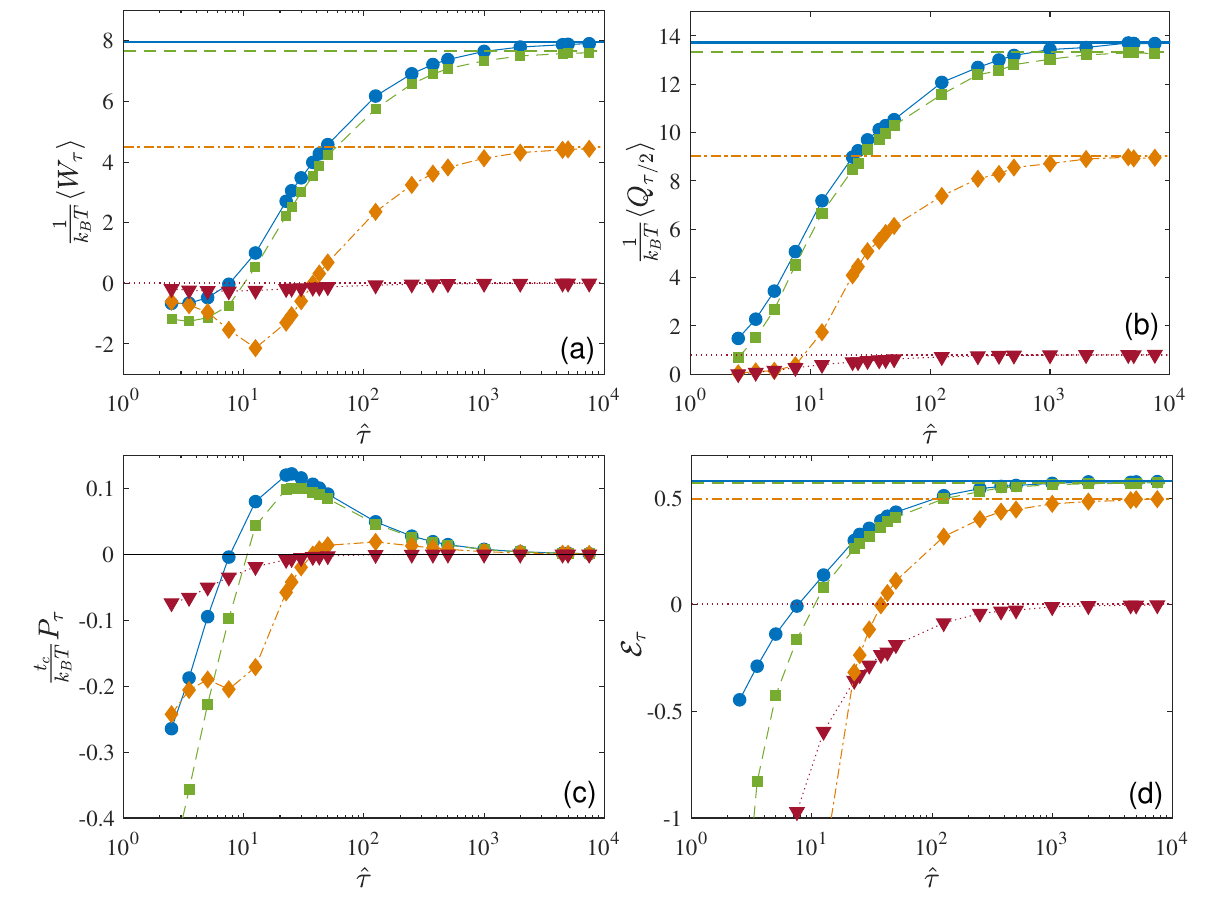}
\caption{Dependence of the different quantities characterizing the finite-time operation of the Stirling engine on the cycle duration normalized by the time-scale $\tau_c$ defined in \eref{eq:tc}. The distinct curves were determined from the numerical simulations of the stochastic differential equations \eref{eq:LangevinBrownianengine} and \eref{eq:AOUvel} with fixed $\alpha = 0.5$, $\theta = 5$, and $\delta = 0.1$: (a) mean output work; (b) mean input heat; (c) mean output power; (d) efficiency. In all subfigures, the symbols correspond to the following pairs of values $(\nu,\epsilon)$ depicted along the identity line $\epsilon = \nu$ in \fref{fig:2}(c): {\bf{A}}$(0.025,0.025)$ ($\fullcircle$, blue solid line), {\bf{B}}$(0.1,0.1)$ ($\fullsquare$, green dashed line), {\bf{C}}$(1,1)$ ($\fulldiamond$, orange dotted-dashed line), and {\bf{D}}$(100,100)$ ($\fulltriangledown$, red dotted line). The horizontal lines in (a), (b) and (d) with the same styles as those connecting the symbols outline the corresponding quasistatic values. The horizontal black solid line in (c) represents the zero-power curve $P_{\tau} = 0$.}
 \label{fig:3}
\end{figure}

\subsection{Finite-time performance}\label{subsect:finitetime}
We now analyze the operation of the Stirling engine for finite cycle durations $\tau < \infty$. For the sake of convenience, we nondimensionalize all lengths and times by means of the characteristic scales
\begin{equation}\label{eq:xc}
    x_c = \sqrt{\frac{k_B T}{\kappa_M}},
\end{equation}
\begin{equation}\label{eq:tc}
    t_c = \frac{\gamma}{\kappa_M},
\end{equation}
respectively, which represent the typical distance that the engine particle travels by thermal agitation inside the harmonic potential at maximum stiffness and its corresponding viscous relaxation time, both in the absence of coupling with the active particle. In order to simulate the fluctuating self-propulsion speed $v(t)$ with the properties described in \eref{eq:meannoise}-\eref{eq:autocorrelnoise}, we introduce the auxiliary stochastic equation of motion for an Ornstein-Uhlenbeck process
\begin{equation}\label{eq:AOUvel}
    \frac{d}{dt}v(t) = -\frac{1}{\tau_r} v(t) + \zeta_v(t),
\end{equation}
with initial condition $v(0) = 0$, where $\zeta_v(t)$ is a white noise term of zero mean, $\langle \zeta_v(t) \rangle = 0$, with autocorrelation function
\begin{equation}\label{eq:autocorrspeed}
    \langle \zeta_v(t) \zeta_v(t')\rangle = \frac{2}{\tau_r} \sigma_v(t) \sigma_v(t') \delta(t - t'),
\end{equation}
and fully decorrelated from the thermal noises $\zeta_X(t)$ and $\zeta_x(t)$.

\begin{figure}[htb]
 \centering
\includegraphics[width=0.95\columnwidth]{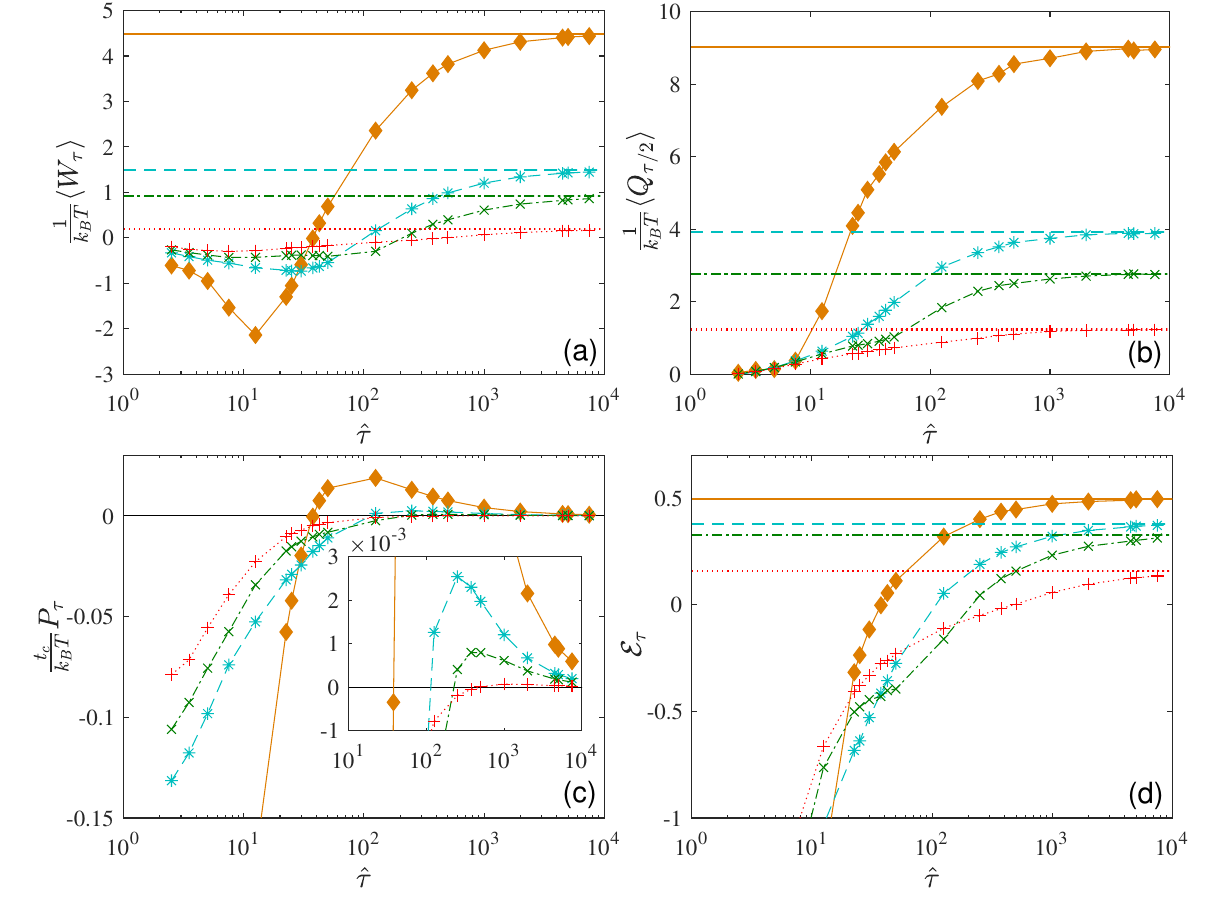}
\caption{Dependence of the different quantities characterizing the finite-time operation of the Stirling engine on the cycle duration normalized by the time-scale $\tau_c$ defined in \eref{eq:tc}. The distinct curves were determined from the numerical simulations of the stochastic differential equations \eref{eq:LangevinBrownianengine} and \eref{eq:AOUvel} with fixed $\alpha = 0.5$, $\theta = 5$, and $\delta = 0.1$: (a) mean output work; (b) mean input heat; (c) mean output power; (d) efficiency. In all subfigures, the symbols correspond to the following pairs of values $(\nu,\epsilon)$ depicted in the region in \fref{fig:2}(c) giving rise to the suboptimal regime: {\bf{C}}$(1,1)$ ($\fulldiamond$, orange solid line), {\bf{E}}$(1,10)$ ($\ast$, turquoise dashed line), {\bf{G}}$(10,1)$ ($\times$, green dotted-dashed line), and {\bf{F}}$(1,100)$ ($+$, red dotted line). The horizontal lines in (a), (b) and (d) with the same styles as those connecting the symbols outline the corresponding quasistatic values. The horizontal black solid line in (c) represents the zero-power curve $P_{\tau} = 0$. The inset in (c) is an expanded view of the main figure.}
 \label{fig:4}
\end{figure}

We simulated numerically the dimensionless versions of \eref{eq:LangevinBrownianengine} and \eref{eq:AOUvel}, see further details in \ref{app}, for fixed values of the parameters $\alpha = 0.5$ (engine and active particles with the same friction), $\theta = 5$, and $\delta = 0.1$, and varying values of the nondimensionalized cycle time, $\hat{\tau} = \tau/\tau_c$, namely $\hat{\tau} = 2.5, 3.5, 5, 7.5, 12.5, 22.5, 25, 30, 37.5, 42.5, 50, 125,
250, 375, 500, 1000, 2000, 4500$, 5000, and 7500. These span more than three orders of magnitude of the shortest viscous relaxation time of the engine particle alone. Having fixed the previous parameters, thirteen pairs of values of $(\nu, \epsilon)$ were chosen for the finite-time simulations as indicated by the different symbols in \fref{fig:2}(b): {\bf{A}}$(0.025, 0.025)$, {\bf{B}}$(0.1, 0.1)$, {\bf{C}}$(1, 1)$, {\bf{D}}$(100, 100)$, {\bf{E}}$(1, 10)$, {{\bf{F}}$(1, 100)$, {\bf{G}}$(10, 1)$, {\bf{H}}$(0.01, 100)$, {\bf{I}}$(0.05, 20)$, {\bf{J}}$(0.1, 10)$,
{\bf{K}}$(1, 0.01)$, {\bf{L}}$(1, 0.03)$, and {\bf{M}}$(1, 0.1)$.

\begin{figure}[htb]
 \centering
\includegraphics[width=0.95\columnwidth]{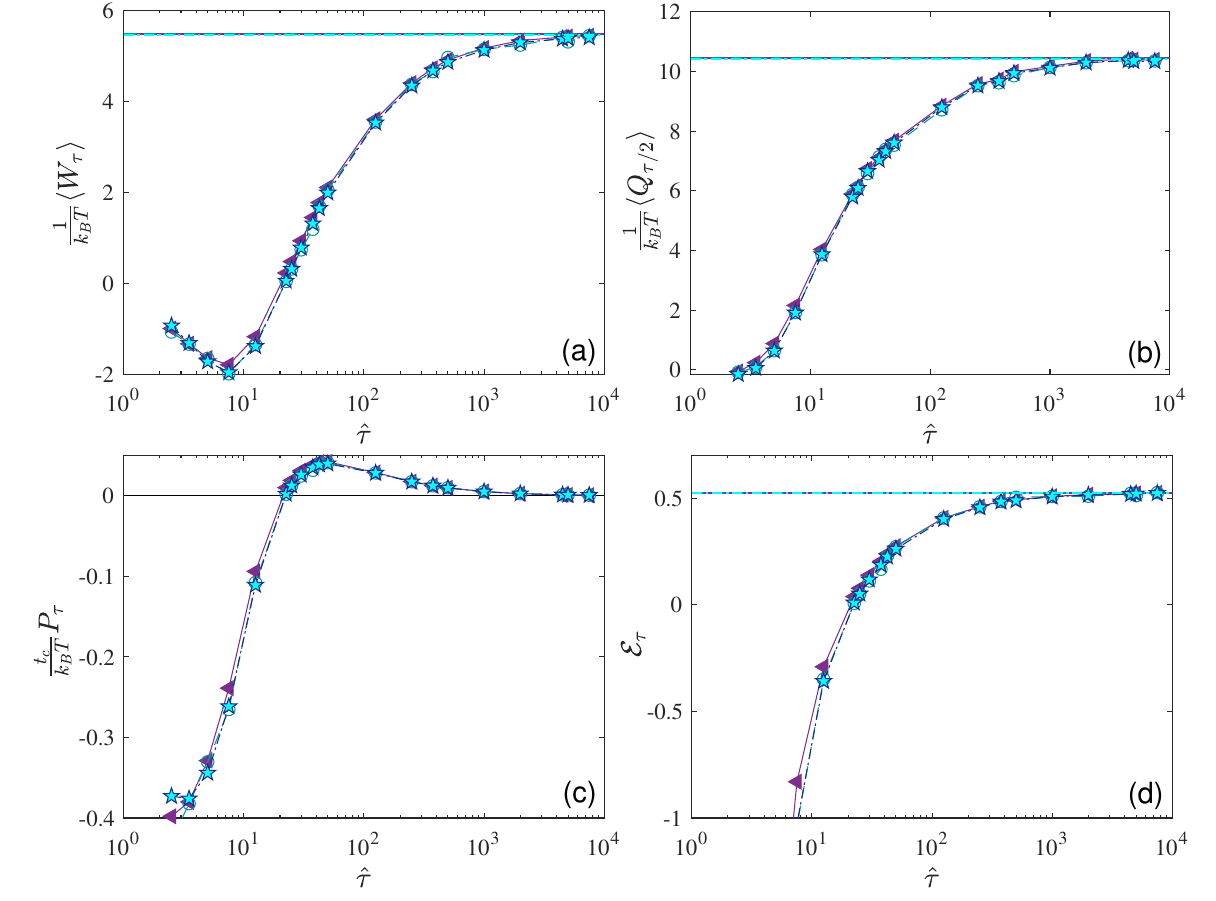}
\caption{Dependence of the different quantities characterizing the finite-time operation of the Stirling engine on the cycle duration normalized by the time-scale $\tau_c$ defined in \eref{eq:tc}. The distinct curves were determined from the numerical simulations of the stochastic differential equations \eref{eq:LangevinBrownianengine} and \eref{eq:AOUvel} with fixed $\alpha = 0.5$, $\theta = 5$, and $\delta = 0.1$: (a) mean output work; (b) mean input heat; (c) mean output power; (d) efficiency. In all subfigures, the symbols correspond to the following pairs of values $(\nu,\epsilon)$ depicted along the curve $\epsilon = 1/\nu$ in \fref{fig:2}(c): {\bf{H}}$(0.01,100)$ ($\blacktriangleleft$, purple solid line), {\bf{I}}$(0.05,20)$ ($\opencircle$, teal dashed line), and {\bf{J}}$(0.1,10)$ ($\bigstar$, cyan dotted-dashed line). The horizontal lines in (a), (b) and (d) with the same styles as those connecting the symbols outline the corresponding quasistatic values. The horizontal black solid line in (c) represents the zero-power curve $P_{\tau} = 0$.}
 \label{fig:5}
\end{figure}

\begin{figure}[htb]
 \centering
\includegraphics[width=0.95\columnwidth]{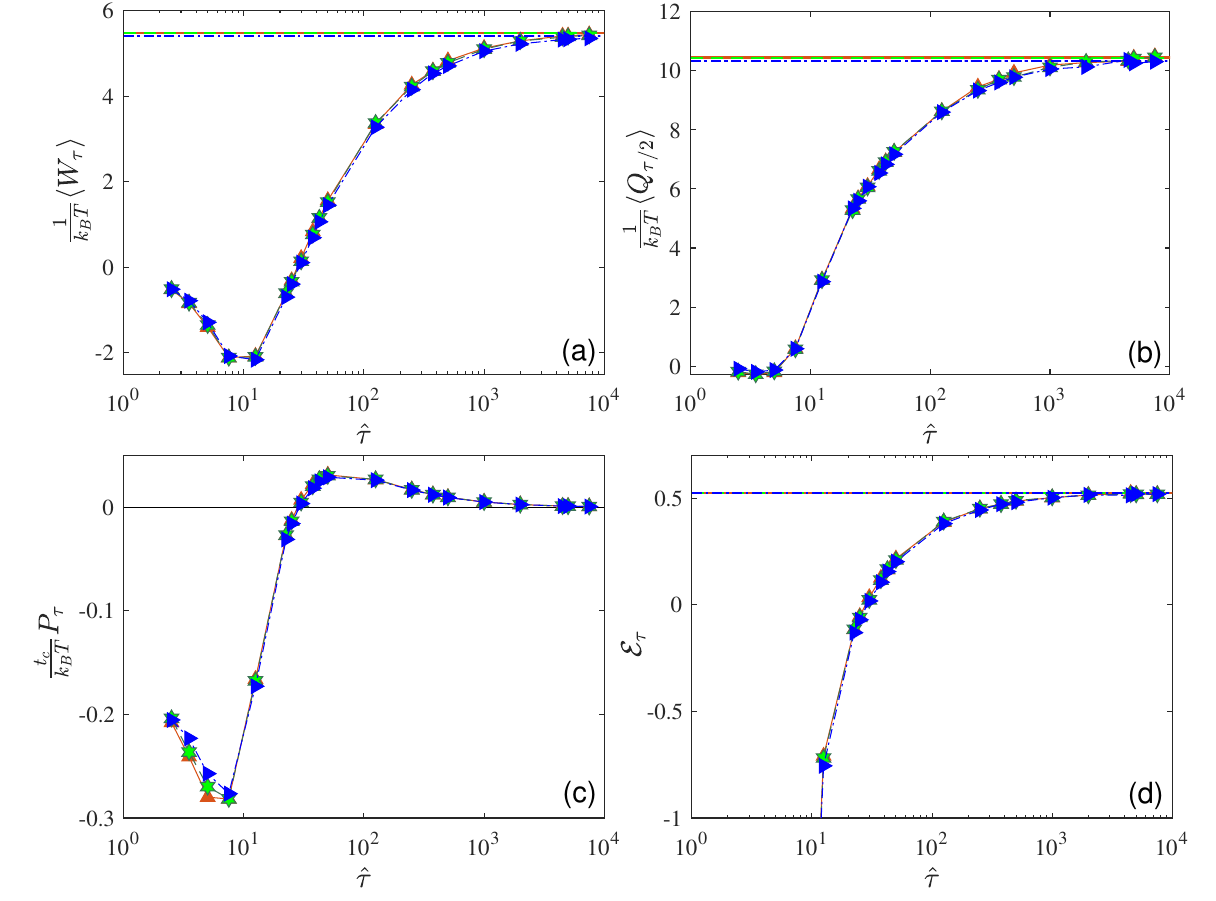}
\caption{Dependence of the different quantities characterizing the finite-time operation of the Stirling engine on the cycle duration normalized by the time-scale $\tau_c$ defined in \eref{eq:tc}. The distinct curves were determined from the numerical simulations of the stochastic differential equations \eref{eq:LangevinBrownianengine} and \eref{eq:AOUvel} with fixed $\alpha = 0.5$, $\theta = 5$, and $\delta = 0.1$: (a) mean output work; (b) mean input heat; (c) mean output power; (d) efficiency. In all subfigures, the symbols correspond to the following pairs of values $(\nu,\epsilon)$ depicted along the line $\nu = 1$ in \fref{fig:2}(c): {\bf{K}}$(1,0.01)$ ($\fulltriangle$, orange solid line), {\bf{L}}$(1,0.03)$ (\ding{86}, green dashed line), and {\bf{M}}$(1,0.1)$ ($\blacktriangleright$, blue dotted-dashed line). The horizontal lines in (a), (b) and (d) with the same styles as those connecting the symbols outline the corresponding quasistatic values. The horizontal black solid line in (c) represents the zero-power curve $P_{\tau} = 0$.}
 \label{fig:6}
\end{figure}

\Fref{fig:3}(a) and \fref{fig:3}(b) show some of the resulting numerical curves that portray the general behavior of the mean produced work and the mean absorbed heat as functions of the cycle duration, respectively. In these figures, four pairs of values $(\nu,\epsilon)$, namely {\bf{A}}$(0.025, 0.025)$, {\bf{B}}$(0.1, 0.1)$, {\bf{C}}$(1, 1)$, and {\bf{D}}$(100, 100)$, were selected along the identity line $\epsilon = \nu$ in the colormap of \fref{fig:2}(c) in order to explore the finite cycle-time behavior over the two different operation regimes identified in the quasistatic limit. As can be clearly seen, for sufficiently large $\hat{\tau}$ all curves converge to their quasistatic values given by \eref{eq:meanWork} and \eref{eq:meanHeat}, where {\bf{A}} and {\bf{B}} (optimal regime) almost reach their highest possible values given by \eref{eq:maxworkheat}, whereas {\bf{C}} (suboptimal regime) exhibits a significant decrease in both quantities, and {\bf{D}} (suboptimal regime) has a vanishingly small mean output work and the minimum mean heat absorption described by \eref{eq:minworkheat}. 

We point out that, while the mean input heat is always positive and increases monotonically with $\hat{\tau}$ until it levels off at its quasistatic value, the mean work displays a non-monotonic dependence on $\hat{\tau}$ and can become negative for sufficiently short cycle durations, thus behaving like a heat pump that consumes work rather than delivering it. This effect, which is also observed in other types of stochastic engines operating at finite cycle durations, originates from the large amount of energy that is irreversibly dissipated when the particle is quickly driven by the cyclic variation of $\kappa(t)$ and $\sigma_v(t)$. Consequently, there exists a particular cycle duration that depends on the specific values of the parameters $\epsilon$ and $\nu$ at which the engine stalls ($\langle W_{\tau} \rangle = 0$), and separates the operation regime as a heat pump ($\langle W_{\tau} \rangle < 0$) from that as a motor ($\langle W_{\tau} \rangle > 0$). Moreover, by taking the ratio between the mean work and the cycle duration, we can determine the corresponding average output power per cycle
\begin{equation}\label{eq:power}
    P_{\tau} = \frac{\langle W_{\tau} \rangle}{\tau}, 
\end{equation}
which is plotted in \fref{fig:3}(c) for the same curves represented in \fref{fig:3}(a). It is noteworthy that, similar to the performance of other heat engines, $P_{\tau}$ exhibits a maximum that originates from the trade-off between the large dissipated energy at short cycles and the slow operation with low dissipation at long cycles. The location of the maximum in $P_{\tau}$ and its magnitude depend on the specific values of the nondimensional parameters of the system in an intricate manner. Nevertheless, the general trend is that, starting from the optimal regime with the largest maximum of the mean power output that takes place at the shortest cycle time, the location of the maximum is shifted to longer and longer cycle durations while its magnitude becomes systematically smaller as the corresponding  quasistatic mean output work decreases when moving toward the suboptimal regime. This observation reveals that the optimal regime not only corresponds to the parameters leading to the highest quasistatic efficiency at fixed $\theta$ and $\delta$ that is given by \eref{eq:maxquaseff} but also provides the broadest range of finite cycle times over which the engine can operate as a motor with the highest output powers. This behavior as a motor also translates into non-negative efficiency curves of the engine operating at finite time that 
monotonically grow from 0 to the quasistatic limit given by \eref{eq:quasistaticefficiency} as $\hat{\tau}$ increases.
For shorter cycles, in all cases the efficiencies become negative when the engine functions as a heat pump consuming work on average. The particular cycle duration at which the efficiency transits from negative to non-negative values becomes larger and larger when moving from the optimal to the suboptimal regime, as demonstrated in \fref{fig:3}(d).

As illustrated in \fref{fig:4} for the data corresponding to {\bf{C}}$(1,1)$, {\bf{E}}$(1,10)$, {\bf{G}}$(10,1)$, and {\bf{F}}$(1,100)$, the finite-time features previously described when exploring the plane $(\nu,\epsilon)$ of its parameter space become very conspicuous in the region of the suboptimal regime close to the curve that separates it from the optimal one represented by the dashed lines in \fref{fig:2}(c). For instance, in \fref{fig:4}(a) we clearly show that a decrease of the quasistatic mean delivered work in this region gives rise to a systematic drop of the average output work at finite operation times with a resulting narrowing of the interval of cycle times over which the engine functions as a motor. This also leads to a progressive reduction of the mean output power, whose maximum diminishes in magnitude and occurs at larger and larger cycle times, as shown in \fref{fig:4}(c). In combination with the monotonic growth with $\hat{\tau}$ of the mean input heat represented in \fref{fig:4}(b), this leads to finite-time efficiencies that change from negative to positive values at specific cycle times that become increasingly shorter as the corresponding quasistatic mean work rises; see \fref{fig:4}(d).

Another relevant aspect is the suboptimal finite-time performance of the Stirling engine for distinct pairs of values $(\epsilon,\nu)$ giving rise to the same quasistatic efficiency along the dashed curve of \fref{fig:2}(c), like those presented in \fref{fig:5} satisfying $\epsilon = 1/\nu$ and those in \fref{fig:6} where $\nu = 1$, both datasets with $\mathcal{E}_{\tau \rightarrow \infty} \approx 0.52$. Interestingly, we find that all the quantities that characterize the engine performance in each separate figure, namely the mean output work, the mean input heat, the mean output power and the efficiency, exhibit the same dependence on $\hat{\tau}$ independently of the specific values of the parameters $\nu$ and $\epsilon$, when working both as a motor or a heat pump. Note that in \fref{fig:5}, the condition $\epsilon = 1/\nu$ means that $\tau_A = \tau_{\kappa_M}$, with $\tau_r = 0.01 \tau_{\kappa_M}$, $\tau_r = 0.05 \tau_{\kappa_M}$, and $\tau_r = 0.1 \tau_{\kappa_M}$ for the data corresponding to {\bf{H}}(0.01,100), {\bf{I}}(0.05,20), and {\bf{J}}(0.1,10), respectively. On the other hand, the condition $\nu = 1$ that holds in \fref{fig:6} requires that $\tau_r = \tau_{\kappa_M}$, with $\tau_A = 0.01 \tau_{\kappa_M}$, $\tau_A = 0.03 \tau_{\kappa_M}$, and $\tau_A = 0.1 \tau_{\kappa_M}$ for the data corresponding to {\bf{K}}(1,0.01), {\bf{L}}(1,0.03), and {\bf{M}}(1,0.1), respectively. In particular, this remarkable result shows that, when either the active persistence time $\tau_r$ or interaction relaxation time $\tau_A$ is comparable to the viscous dissipation time $\tau_{\kappa_M}$, the engine exhibits a finite-time performance with positive output power and efficiency that become insensitive to the remaining time-scale provided that this is shorter than $\tau_{\kappa_M}$.

\section{Conclusion}\label{sec:conc}

In this paper, we have examined a minimal model of an active Brownian engine powered by Stirling-like cycles applied by time-periodic variations of a harmonic potential and the activity of a self-propelled particle that is elastically coupled to the working substance, both in contact with a thermal bath at constant temperature. The engine performance is determined by the relative magnitudes of the persistence time of the fluctuating speed of the active particle, its relaxation time due to the interaction with the engine particle, and the viscous relaxation time of the latter in the confining potential. In the quasistatic limit, we could derive analytic formulae for the mean output work, the mean input heat, and the corresponding efficiency, which could be expressed in terms of ratios of such three time-scales as well as other dimensionless parameters describing characteristics of the cycle and the environment. This analysis has allowed us to clearly identify two regimes of the quasistatic engine performance: an optimal one with a high mean output work and an efficiency very close to an upper bound,  which occurs only when the viscous relaxation time of the working substance is the dominant time-scale, and a suboptimal one displaying a hindered mean output work and efficiency when this condition is not satisfied. In addition, by numerically investigating the non-quasistatic behavior of the different quantities that characterize the engine performance, we have verified that such distinctive properties of the quasistatic case persist at finite times for sufficiently large cycle durations. For shorter cycles, the engine exhibits a positive efficiency that is generally lower than the quasistatic one but with a positive mean output power, whose maximum depends on the specific values of the system parameters. The broadest intervals of cycle times for positive power production and the highest output powers occur for parameters corresponding to those at which the optimal quasistatic regime is achieved, whereas diminished positive powers over narrower cycle-time ranges are obtained for parameters coinciding with the suboptimal regime. In all cases, for sufficiently short cycle times, the engine behaves as a heat pump rather than a motor, where the mean output work, power, and efficiency are negative. 

The model studied in this paper represents a simplified version of the operation of a cyclic colloidal engine in an active viscoelastic fluid, where the mechanical coupling between the engine and the environment generally gives rise to memory effects in the dynamics of the former. Therefore, we hope that the results presented here will trigger further efforts to investigate other properties of stochastic engines in non-Markovian baths such as those commonly encountered in biological systems. Moreover, whether a meaningful effective temperature can be defined to map the dynamics of the
engine studied here into that of a single particle in an effective equilibrium bath obeying second-law bounds like in \cite{holubec2020} is an interesting question that will be the subject of future research.

\ack

We acknowledge support from DGAPA-UNAM PAPIIT Grant No. IA104922.

\appendix

\section{Numerical simulation of finite-time operation}\label{app}

To carry out the finite-time simulations presented in subsection \ref{subsect:finitetime}, equations \eref{eq:LangevinBrownianengine} and \eref{eq:AOUvel} were numerically solved by means of the Euler–Maruyama method, whose discretized versions, upon normalizing by the characteristic scales defined in \eref{eq:xc} and \eref{eq:tc}, read as
\begin{eqnarray}\label{eq:numerical}
    \hat{X}_{i+1} & = & \left[1 - \left( \hat{\kappa}_i + \frac{1 - \alpha}{\epsilon \nu}\right)\Delta \hat{t} \right] \hat{X}_i + \frac{1-\alpha}{\epsilon \nu} \Delta \hat{t} \hat{x}_i + \sqrt{2 \Delta \hat{t}} N_X, \nonumber\\
 \hat{x}_{i+1} & = & \left( 1 - \frac{\alpha}{\epsilon \nu} \Delta \hat{t}\right) \hat{x}_i + \frac{\alpha}{\epsilon \nu} \Delta \hat{t} \hat{X}_i + \Delta \hat{t} \hat{v}_i + \sqrt{\frac{2\alpha \Delta \hat{t}}{1-\alpha}} N_x, \nonumber\\
    \hat{v}_{i+1} & = & \left(1 - \frac{\alpha}{\nu} \Delta \hat{t} \right) \hat{v}_i + \sqrt{\frac{2 \alpha^3 \Delta \hat{t}}{\delta (1 - \alpha)^2 \nu^2 }} N_v.
\end{eqnarray}
In \eref{eq:numerical}, $N_X$, $N_x$, and $N_v$ are independent random numbers drawn from the standard normal distribution at each iteration, and $\hat{\kappa}_i$ is the value of the stiffness $\kappa$ (normalized by $\kappa_M$) at the $i-$th step of the iteration according to \eref{eq:Stirlingkappa}. Here, the $i-$th step corresponds to the discretized dimensionless time $\hat{t} = i \Delta \hat{t}$ with time-step size $\Delta \hat{t} = 0.001$, where $i$ is an index that takes the values $i = 0, \ldots, M$, and $M$ is an integer chosen so that 300 full cycles are simulated for a given initial condition $\hat{X}_0$, $\hat{x}_0$, $\hat{v}_0$. For example, for the shortest time interval simulated ($\hat{\tau} = 2.5$), $M = 7.5\times 10^5$, while for the largest one ($\hat{\tau} = 7500$), $M = 2.25\times 10^9$. The initial conditions of a single trajectory that spans 300 cycles are selected such that $X_0$ is a random number drawn from the standard normal distribution, whereas we set $\hat{x}_0 = 0$, $\hat{v}_0 = 0$. In addition, 100 distinct trajectories corresponding to 100 independent random initial values of $\hat{X}_0$ were also simulated to increase by two orders of magnitude the number of cycles to calculate the necessary average quantities. We verified that after the first 10 cycles, which are later discarded from the analysis, the probability distribution of $\hat{X}$ reaches a periodic steady state. Therefore, all the mean quantities reported in the paper characterizing the finite-time behavior of the engine result from averaging over $290\times 100 = 2.9 \times 10^4$ cycles.

\end{document}